# Component Interaction Graph: A new approach to test component composition

Arup Abhinna Acharya and Sisir Kumar Jena

**Abstract**—The key factor of component based software development is component composition technology. A Component interaction graph is used to describe the interrelation of components. Drawing a complete component interaction graph (CIG) provides an objective basis and technical means for making the testing outline .Although many researches have focused on this subject, the quality of system that is composed of components has not been guaranteed. In this paper, a CIG is constructed from a state chart diagram and new test cases are generated to test the component composition.

———————————— ◆ ————————————

## 1 INTRODUCTION

Component-based software engineering (CBSE) is increasingly being adopted for software development. This approach, which uses reusable components as building blocks for constructing software, can facilitate fast-paced delivery of scalable, evolvable software systems. It also offers the promise of putting the improvement of software quality and productivity into common practice [1]. A component-based software system often consists of a set of self-contained and loosely coupled components allowing plug-and-play. The components may have been written in different programming languages, executed in various operational platforms, and distributed across vast geographic distances; some components may be developed in-house, while others may be third party or commercial off-the-shelf components whose source code may not be available to developers. These component-based software characteristics facilitate development of complex systems by allowing integration of reusable components from different vendors and by providing flexibility for dynamic software evolution.

The paper is organized as follows: Section 2 describes about the basic concept and introduces the concept of CIG. This section also describe about a state chart diagram. Generation of CIG from a state chart diagram is presented in section 3. We end this paper with our conclusion and future research in section 4.

## 2 BASIC CONCEPTS

Component interfaces define all content of interaction with the external and it is the only way that a component communicates with the outside. Component composition can be defined by interface to express the ideal condition, thus we regard the component as a union of interfaces.

- *Arup Abhinna Acharya is with the School of Computer Engineering, KIIT University, Bhubaneswar.*
- *Sisir Kumar Jena is with the Department of CSE, Nalanda Institute of Technology, Bhubaneswar.*

There are two kinds of interface: service providing and service required while the services are provided by an environment or other components. In our approach, all components are considered plug-compatible in the sense that a service required interface can be connected to a service providing interface. For simplification, we take the environment as an abstract component [2].

Component $C = (P, R)$, where
$P = \{P_1, P_2 \ldots, P_n\}$ is the set of providing services interface,

$R = \{R_1, R_2, \ldots, R_m\}$ is the set of required services interface.

The providing and required services of a component C is denoted by C.P and C.R respectively and $C.P \cap C.R = \Phi$

There is an interface mapping from required services to providing services of a component that is $f : C.R \rightarrow C.P$. Test case could be produced after the analysis of function f according to the specification of component interfaces [2].

The composition of two component means that the required services of one component are provided by another partly or fully.

let $C_1 = (P_1, R_1)$, $C_2 = (P_2, R_2)$,
if $C_1.P_1 \cap C_2.R_2 \neq \Phi$ or $C_2.P_2 \cap C_1.R_1 \neq \Phi$ then $C_1$ and $C_2$ is composable.

Let S be the set of satisfied services, $S = (C_1.P_1 \cap C_2.R_2) \cup (C_2.P_2 \cap C_1.R_1)$, the composition of $C_1$ with $C_2$ is denoted by $C_1 \circledP C_2$.

$C_1 \circledP C_2 = (P, R)$ where $P = C_1.P_1 \cup C_2.P_2 / S$, $R = C_1.R_1 \cup C_2.R_2 / S$.

The providing services of $C_1 \circledP C_2$ are the union of $C_1.P_1$ and $C_2.P_2$ with the remove of consumed services for S.

The required services of C1 ℗ C2 are the union of C1.R1 and C2.R2 with the remove of satisfied services in S. With the definition of composition the providing and required services are propagated to the interface of composed component, so the composition could be carried parallel.

After the composition a new test case set for the composed component *Tnew* could be created. Let T1 is the test case for C1 and T2 for C2, then T C1 ℗ C2= ((T1UT2)/Ts) U *Tnew*, where Ts are the test cases related to satisfied services in T1 and T2, Tnew are the produced test cases after composition, that is to say, the test cases for composed component are the union of unaffected part of originally test cases and new produced test cases for the matching of required and providing services.

A Component interaction graph is used to describe the interrelation of components. Drawing a complete component interaction graph (CIG) provides an objective basis and technical means for making the testing outline. Generally speaking, a CIG can be obtained from analysis of the requirement and design documents of a system. We simplified the CIG model introduced in reference [1] to describe the interaction of components.

A CIG is a directed graph where CIG = (V, E), V = VI U VC is a set of nodes, VI is the set of interface nodes and VC is the set of component nodes, E represents the set of directed edges. If there is an existing edge form C1.P1 to C2.R1 in the CIG it means the required service R1 of C2 has been satisfied by the providing service P1 of C1, which is C2.R1= C1.P1. In the bottom-up development if all required services of a component are satisfied, then it can be assembled into the realized portion.

We denote an interface with an ellipse and a component with dashed square; the interfaces belong to one component that was drawn in the same square. The interaction among components can be gained from the CIG directly.

There are two kinds of special components, one is the component without the required services, and can be the beginning of composition, the other is the one without the providing services for other components, and can be one end of the composition. According to the fact the numbers of two kinds of components can be one or more.

The CIG illustration is given below:

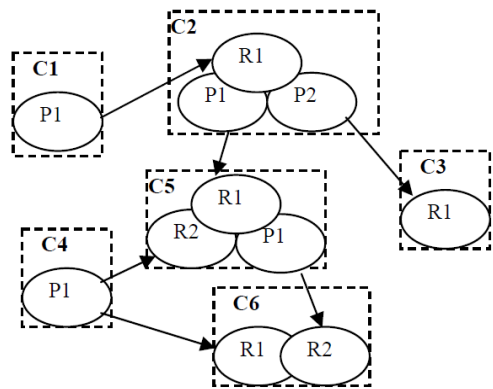

Figure 2.1: Component Interaction Graph

State chart diagrams in UML are used to model the dynamic aspects of a system. A state chart diagram consists of states, transitions, events and actions [Booch, Rumbaugh and Jacobson 1998] and it shows a state machine emphasizing the flow of control from state to state.

## 3 GENERATION OF CIG

We model component interactions using a Component Interaction Graph (CIG) which depicts interaction scenarios among components [1]. A component can be represented in the form of a state chart diagram. Here we present a way to convert a statechart diagram to a component interaction graph.

A statechart diagram can be defined as a collection of different state of a component:
　　S={ s1, s2, …, sn}
If the machine receives an event while in state $s_i$, it reaches state $s_j$, from which an additional event takes it to state $s_k$. Here we can say $s_i$ is a provided service for $s_j$ and in reverse $s_j$ requires services from $s_i$. At this situation we search the state which triggers an event so that the other component activates and reach at its initial state. Accordingly the state of the second component is a required state or required interface for that component.

To further explain the above concept, let a component C1 has a state S1 and a component C2 has a state S2. Now S2 is said to be a required interface if trigger of an event at S1 automatically activate C2 and reach at a initial state S2.

**Algorithm:**
Precondition: test case library T = {T1, T2….,Tn)
Component Library C={C1, C2, …, Ck}
State S={S1, S2, …, Si}
Output: CIG
1. Start
2. Remove the state which switches between the components.
3. Repeat for (k=1; k < M; k++)
　　Repeat for (i=1; i<N; i++)
　　　　a. Search a state Si from the component Ck from which, occurring of an event acti-

        vate a component Cj, where j=1, 2, ..., Q.
      b. Set Si as Pi // Where P is stand for provided interface (state)
    End loop
  End loop
4. Repeat for (j=1; j<Q ; j++)
    Repeat for (i=1; i<N; i++)
      a. Search a state Si from the component Cj which requires services.
      b. Set Si as Ri // Where R stand for required interface (state)
    End loop
  End loop

5. If the state is not a Pi || Ri Set it as Gi. // Gi stands for a intermediate state or a initial/end state that requires for test case generation.
6. End

### 3.1 Example: Vending Machine

The example represents a case in which component A uses a component C. The application models a vending machine. A user can insert coins into the machine, ask the machine to cancel the transaction, which results in the machine returning all the coins inserted and not consumed, or ask the machine to vend an item. If an item is not available, a user's credit is insufficient, or a selection is invalid, the machine prints an error message and does not dispense the item, but instead returns any accumulated coins.

Figure 3.1 shows a component diagram that represents application VendingMachine, component Dispenser, and their interactions. Dispenser provides an interface that is used by VendingMachine. VendingMachine uses the services provided by Dispenser to manage credits inserted into the vending machine, validate selections, and check for availability of requested items.

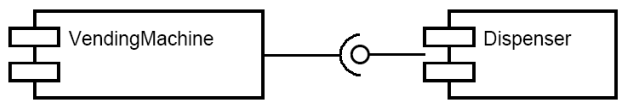

Figure 3.1: Component diagram for vending machine and dispenser.

Figure 3.2 shows a statechart specification of VendingMachine. The machine has five states (NoCoins, SingleCoin, MultipleCoins, ReadyToDispense, and Dispensing), accepts five events (insert, cancel, vend, nok, and ok), and produces three actions (setCredit, dispense, and returnCoins). If the machine receives an insert event while in state NoCoins, it reaches state SingleCoin, from which an additional insert event takes it to state MultipleCoins; if the machine receives a vend or cancel event while in state NoCoins, it remains in that state. In states SingleCoin and MultipleCoins, a vend event triggers action setCredit (with different parameters in the two cases: 1 and 2::max, respectively) and brings the machine to state ReadyToDispense. (The notation 2::max is used to indicate that the value of the parameter can vary between 2 and some predefined maximum.) In state ReadyToDispense, the machine produces action dispense and enters state Dispensing, from which it returns to state NoCoins when it receives a nok or ok event; in both cases, any remaining coins are returned through a returnCoins event.

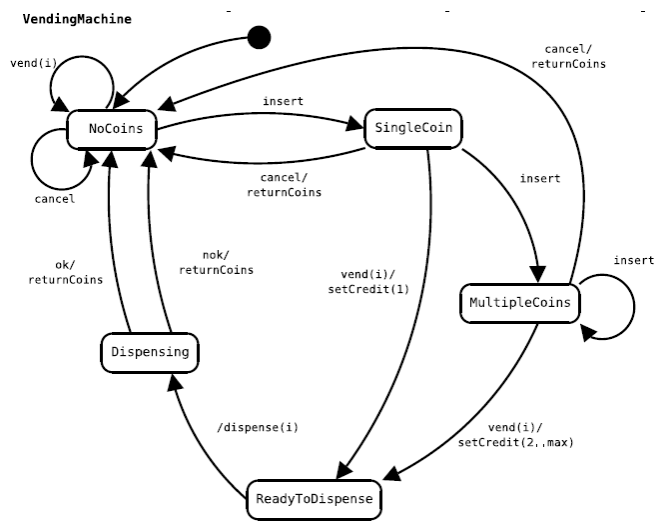

Figure 3.2: State chart specification for vending machine

Figure 3.3 shows a statechart speci_cation of Dispenser. The machine has three states (Empty, Insufficient, and Enabled), accepts two events (setCredit and dispense), and produces two actions (nok and ok). In state Empty, the machine accepts event setCredit and stays in state Empty, reaches state Insufficient, or reaches state Enabled based on the value of the credit, as speci_ed by the guards in the figure. In all three states, the machine accepts event dispense. Events dispense triggers a nok or ok action depending on the availability of the requested item.

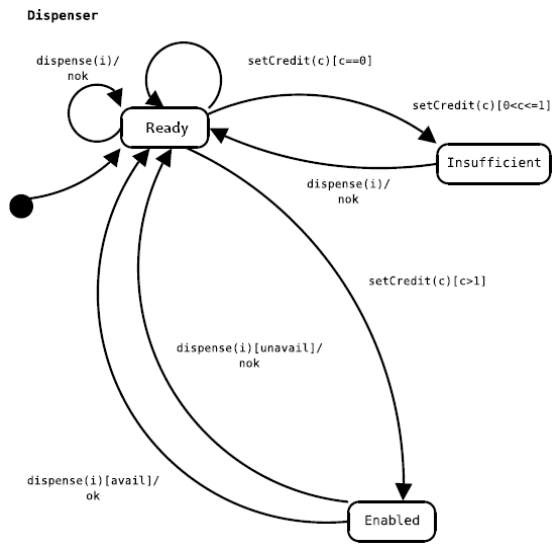

Figure 3.3: State chart specification for Dispenser

the matching of required and providing services.

## 4 CONCLUSION

We have presented a new approach for testing component-based software. Our empirical studies show that testing component-based software is necessary yet expensive. The technique we propose includes several criteria for determining test adequacy. Our on-going research directions on this topic are the development of a tool to support automation of the technique, and enhancement of the technique for automated generation of new test cases without modifying the test case library. The future work is going to explain an algorithm to generate the test cases from the generated CIG.

Now we can generate CIG from the above two state chart diagram by applying the above algorithm. In the component vending machine, the state ready to dispense is switches to the next component Dispenser. So in CIG we can remove the state ready to dispense.

In the following figure 3.4, the state single coin and multiple coin causes the component Dispenser to be activated, so this two states can be act as the provided interface and the ready state of Dispenser is act as the required interface. In the same manner Enabled state of the dispenser is act as the provided interface and Dispensing state is the required interface.

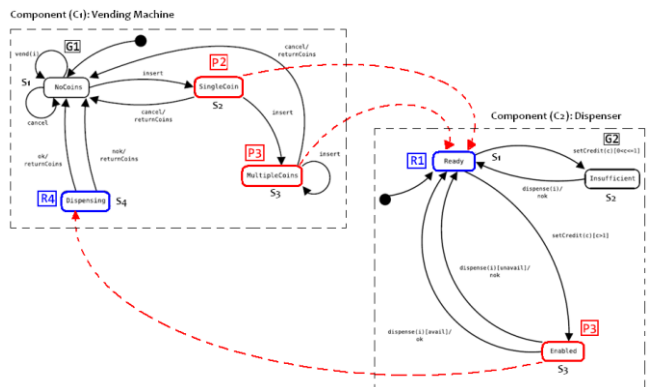

Figure 3.4: CIG generated from state chart diagram

### 3.1 Test case Generation

After the composition a new test case set for the composed component Tnew could be created. Let T1 is the test case for C1 and T2 for C2, then T C1 ⓟ C2= (T1UT2)/Ts)U Tnew, where Ts are the test cases related to satisfied services in T1 and T2, Tnew are the produced test cases after composition, that is to say, the test cases for composed component are the union of unaffected part of originally test cases and new produced test cases for